\documentclass[aps,prl,twocolumn,superscriptaddress]{revtex4}

\usepackage{graphicx}
\usepackage{epsfig}             

\begin{document}


\title{Quantum study of information delay in electromagetically induced transparency}

\author{M.~T.~L.~Hsu}
\affiliation{ARC COE for Quantum-Atom Optics, Australian National University, Canberra, ACT 0200, Australia}

\author{G.~H\'etet}
\affiliation{ARC COE for Quantum-Atom Optics, Australian National University, Canberra, ACT 0200, Australia}

\author{O.~Gl\"ockl}
\affiliation{ARC COE for Quantum-Atom Optics, Australian National University, Canberra, ACT 0200, Australia}

\author{J.~J.~Longdell}
\affiliation{ARC COE for Quantum-Atom Optics, Australian National University, Canberra, ACT 0200, Australia}
\affiliation{Laser Physics Centre, RSPhysSE, Australian National University, Canberra, ACT 0200, Australia}

\author{B.~C.~Buchler}
\affiliation{ARC COE for Quantum-Atom Optics, Australian National University, Canberra, ACT 0200, Australia}

\author{H.-A.~Bachor}
\affiliation{ARC COE for Quantum-Atom Optics, Australian National University, Canberra, ACT 0200, Australia}

\author{P.~K.~Lam}
\email[Email: ]{ping.lam@anu.edu.au}
\affiliation{ARC COE for Quantum-Atom Optics, Australian National University, Canberra, ACT 0200, Australia}

\date{\today}

\begin{abstract}
Using electromagnetically induced transparency (EIT), it is possible to delay and store light in atomic ensembles.  Theoretical modelling and recent experiments have suggested that the EIT storage mechanism can be used as a memory for quantum information. We present experiments that quantify the noise performance of an EIT system for conjugate amplitude and phase quadratures. It is shown that our EIT system adds excess noise to the delayed light that has not hitherto been predicted by published theoretical modelling.  In analogy with other continuous-variable quantum information systems, the performance of our EIT system is characterised in terms of conditional variance and signal transfer.
\end{abstract}

\pacs{42.50.Gy, 03.67.-a}

\maketitle

Following theoretical proposals \cite{fleischhauer}, electromagnetically induced transparency (EIT) \cite{boller} has become the subject of much interest for controlled atomic storage of quantum states of light.  Indeed, the delay and storage of optical qubits in an atomic medium via EIT has recently been shown allowing, in principle, the synchronisation of quantum information processing systems \cite{eisaman, chaneliere}.  Earlier works with classical signals in a vapour cell \cite{kash} and cold atoms \cite{hau} have shown large signal delay with group velocities as low as 17~ms$^{-1}$.  Storage of classical pulses has also been shown for atomic vapour cells \cite{phillips,bajcsy}, cold atomic clouds \cite{liu}, and solid state systems \cite{longdell, turukhin} (although it should be noted that alternative interpretations of such pulse storage experiments have also been published \cite{alexandrov, lezama}).  One experiment \cite{akamatsu} has even shown the transmission of a squeezed state through an EIT system in a vapour cell under the conditions of very small delay. While these experiments are all excellent demonstrations of EIT, to the best of our knowledge, no attempt has been made to quantify the efficacy of EIT for continuous-variable quantum information systems.

Quantum-theoretical treatments of delay and storage via EIT, in the presence of decoherences, have suggested that no excess noise is added to the delayed light \cite{dantan, fleish, matsko, peng}. These works show that the degradation of a quantum state in an EIT system results from - (i) the finite transparency window and (ii) a degradation in the transparency induced by ground state dephasing. The implication is that, within the EIT window and for small ground state dephasing, quantum states of light can be delayed and preserved in an EIT medium. In this letter, we present experimental results that examine the quantum noise performance of an EIT system for conjugate amplitude and phase quadratures, that are measured at sideband frequencies ($\omega$) around the optical carrier.  Since much work on EIT is motivated by quantum information processing, we evaluate the performance of the EIT system using well established criteria for continuous variable (CV) quantum state measurements. In analogy with quantum teleportation and non-demolition experiments where states are transferred from an input to an output, we utilise the conditional variance and signal transfer coefficients to quantify the quantum noise properties of our EIT system. 

\begin{figure}[!ht]
\begin{center}
\includegraphics[width=7.2cm]{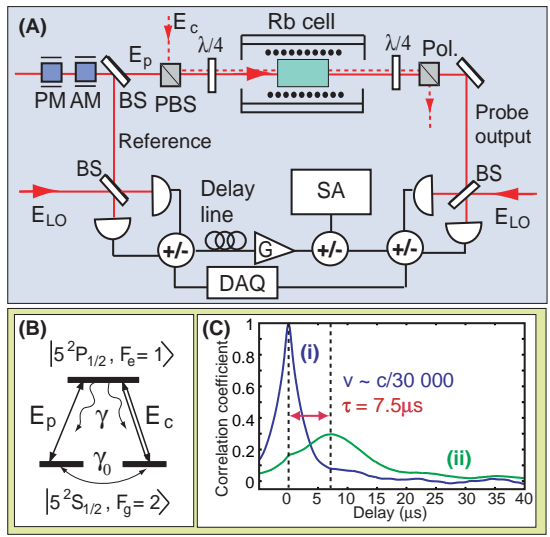}
\caption{(A) Schematic of experimental layout. BS: beamsplitter, PBS: polarising beamsplitter, SA: spectrum analyser, DAQ: data acquisition device, $\lambda/4$: quarter wave-plate, $\lambda/2$: half wave-plate and Pol.: polariser. (B) Atomic level scheme used in our experiment. $E_{p}$ is the probe field, $E_c$ is the pump field, $\gamma$ is the spontaneous emission rate and $\gamma_{0}$ is the ground state dephasing rate. (C) Amplitude quadrature correlation plots. Similar results were observed for the phase quadrature correlation. Cell temperature = 62$^{\circ}$C, probe and pump power densities are 0.32~mW/cm$^2$ and 3.2~mW/cm$^2$, respectively.}
\label{schematic}
\end{center}
\end{figure}

EIT is created by the interaction of probe ($E_{p}$) and pump ($E_{c}$) fields, in a 3-level $\Lambda$-atomic system, as shown in Fig.~\ref{schematic}~(B). When the two fields are resonant with their respective transitions, a quantum interference occurs between the two atomic absorption pathways.  Consequently, a narrow transparency window is generated on resonance \cite{boller}.  Via the  Kramers-Kronig relations, this sharp transparency feature gives rise to strong dispersion and therefore reduced group velocity. Solving the optical Bloch equations for the 3-level atomic system and assuming  $|E_{c}| \gg |E_{p}|$, we find the probe susceptibility to be
\begin{equation} \label{abs}
\chi(\omega) = \frac{i 2 N |g|^{2} (\gamma_{0} - i \omega)}{c k \left( (\gamma_{0} - i \omega)(\gamma - i \omega) + |g E_{c}|^{2} \right)},
\end{equation}
where $k$ is the wave number, $N$ is the atomic density, $g$ is the atom-field coupling constant, $\gamma$ is the spontaneous emission rate and $\gamma_0$ is the ground state dephasing rate.   The imaginary and real parts of $\chi(\omega)$ describe the transmission and dispersion, respectively, of the probe beam.  On propagation through an EIT medium of length $L$, the transmissivity of the probe beam is given by $\eta(\omega) = e^{-\Re\{ ik \chi(\omega) L\} }$. Note that for $\gamma_{0}=0$ and $\omega=0$, the EIT system has perfect transmission.

A schematic of our experiment is shown in Fig.~\ref{schematic}~(A). 
The experiment was driven using a Ti:Sapphire laser, tuned to the $|5^{2} S_{1/2}, F_{g} = 2 \rangle$ to $|5^{2} P_{1/2}, F_{e} = 1 \rangle$ transition of the $D_{1}$ line of Rubidium-87 ($^{87}$Rb). A beam was encoded with sideband amplitude and phase modulation signals. One half of this beam was sent to a homodyne detection system, as a reference beam for the input. The remainder of this beam was used as a probe, by combining with an orthogonally polarised pump beam at a polarising beamsplitter. The overlapping pump and probe beams were converted to left and right circularly polarised modes by a quarter-wave plate before entering an uncoated, isotopically enhanced $^{87}$Rb vapour cell. The heated vapour cell was shielded in two layers of high permeability alloy, that reduced stray magnetic fields to $\le 1$~mG. The probe beam was extracted from the output of the cell using a polariser and sent to a second homodyne detector.  The signals from the two homodyne detectors were monitored using a data acquisition system and electronic spectrum analyser. 

The group velocity in an EIT system is typically quantified by measuring the delay of pulses. In our continuous wave system, we applied a noise modulation (bandwidth of 60~kHz) to our probe beam and then measured the time correlation functions of the probe signals.  By comparing the auto-correlation of the reference beam (Fig.~\ref{schematic}~C(i)) with the cross-correlation of the reference and probe output beams (Fig.~\ref{schematic}~C(ii)) we can accurately measure the delay due to EIT. The maximum of the reference auto-correlation occurred at zero time delay with a width of 60~kHz, which corresponded exactly to the modulation bandwidth of the probe.  The maximum correlation between the reference and probe output beams occurred at a time delay of 7.5~$\mu s$. This corresponded to a group velocity reduction of the input probe beam to $\sim c/30000$. The width of curve (ii) is broader than curve (i), indicating a varying time delay as a function of sideband frequency. Note that a periodic modulation feature was also observed, which was attributed to the resonant modulation locking signals intrinsic to the Ti:Sapphire laser system.

We now analyze the noise performance of our EIT system as a quantum delay line for sidebands at a frequency $\omega$, relative to the carrier frequency of the probe field. Ideally, we would like to possess {\it a priori} information about the probe input state and then use this information to obtain the conditional variance ($V_{\rm in|out}^{\pm}$) between the probe input and output.  Without a pair of entangled beams at our disposal, such a direct measurement of $V_{\rm in|out}^{\pm}$ is not possible. In practice we measure the conditional variance between the probe reference and output beams, from which we can infer $V_{\rm in|out}^{\pm}$ between probe input and output.   The amplitude and phase quadratures are defined in terms of the Fourier transformed annihilation and creation operators, 
as $\hat{X}^{+}(\omega) = \hat{a} (\omega) + \hat{a}^{\dagger} (\omega)$, and $\hat{X}^{-} (\omega) = i (\hat{a}^{\dagger} (\omega) - \hat{a} (\omega))$, respectively.  The conditional variance is measured by  minimising the subtraction of input and output signals with variable gain $G(\omega)$ and time delay $\tau(\omega)$, giving
\begin{equation} \label{vcveit}
V^{\pm}_{\mathrm{in|out}}(\omega)={\rm min}|_{G,\tau} \langle |\hat{X}^{\pm}_{\rm out}(\omega)-G (\omega)e^{i\omega \tau(\omega)} \hat{X}^{\pm}_{\rm in}(\omega) |^2 \rangle. \label{CV}
\end{equation}
For an ideal lossless delay line, the conditional variance limit is given by $V^{\pm}_{\mathrm{in|out}}(\omega)=0$, since the input and output are exactly equal. 
A more practical benchmark is an ideal delay line with some inherent passive loss.  Eq.~(\ref{abs}) shows that EIT has frequency dependant  transmissivity $\eta(\omega)$.  The quantum limit of the conditional variance is therefore found by assuming that this loss is passive, in the sense that transmissivity $\eta(\omega)$ implies the addition of $1-\eta(\omega)$ quanta of vacuum noise. In this case the quantum limit to the conditional variance is $V^{\pm}_{\mathrm{in|out}}=1-\eta(\omega)$. Quantum models have shown that the EIT system should reach this passive loss limit, so that an experiment that compares the conditional variance to this quantum limit is a good test of theory. Experimentally, we find the quantum limit by replacing the gas cell with a beamsplitter that has the same transmissivity as the EIT system. The transmission of the beamsplitter must be changed for each sideband frequency to account for the finite EIT window \footnote{The frequency dependent loss was found by measuring the reduction of signal size for large modulations.}. Since $V_{\rm in|out}^{\pm}$ for a simple passive loss is entirely predictable, this `beamsplitter benchmark' is also a key indicator that our setup correctly measures $V_{\rm in|out}^{\pm}$.

\begin{figure}[!ht]
\begin{center}
\includegraphics[width=7.8cm]{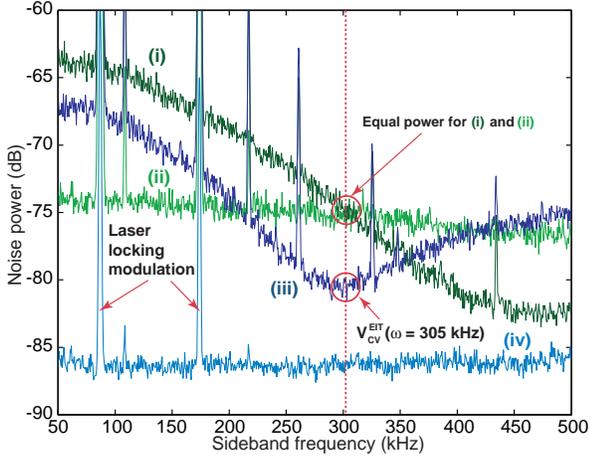}
\caption{$V_{\rm in|out}^{\pm}$ for the amplitude quadrature, optimised for the sideband frequency of 305~kHz. The curves represent the (i) output probe signal, (ii) reference signal with gain $G$ and delay $\tau$, (iii) $V_{\rm in|out}^{\pm}$ between the reference and output signals and (iv) $V_{\rm in|out}^{\pm}$ for the beamsplitter benchmark. The modulation peaks at 87~kHz and 174~kHz are the laser locking signals. Cell temperature = 57$^{\circ}$C, probe and pump power densities are 9.6~mW/cm$^2$ and 96~mW/cm$^2$, respectively .  Measurements were made with a resolution bandwidth (RBW) = 1~kHz, video bandwidth (VBW) = 30~Hz and 5 averages.}
\label{gaincable}
\end{center}
\end{figure}
A sample set of $V_{\rm in|out}^{\pm}$ data is shown in Fig.~\ref{gaincable}. The output (i) and  reference (ii) signals, intersect at a sideband frequency of 305~kHz, corresponding  to the frequency at which the $V_{\rm in|out}^{\pm}$ for the EIT system (iii) is minimum.  The $V_{\rm in|out}^{\pm}$ for the beamsplitter benchmark (iv) is lower than the minimum point of curve (iii), indicating that the delayed probe beam has excess noise. Since the EIT system has frequency dependent absorption and delay, the gain and time delay for the $V_{\rm in|out}^{\pm}$ measurement had to be optimised for each measurement frequency. 

\begin{figure}[!ht]
\begin{center}
\includegraphics[width=7.8cm]{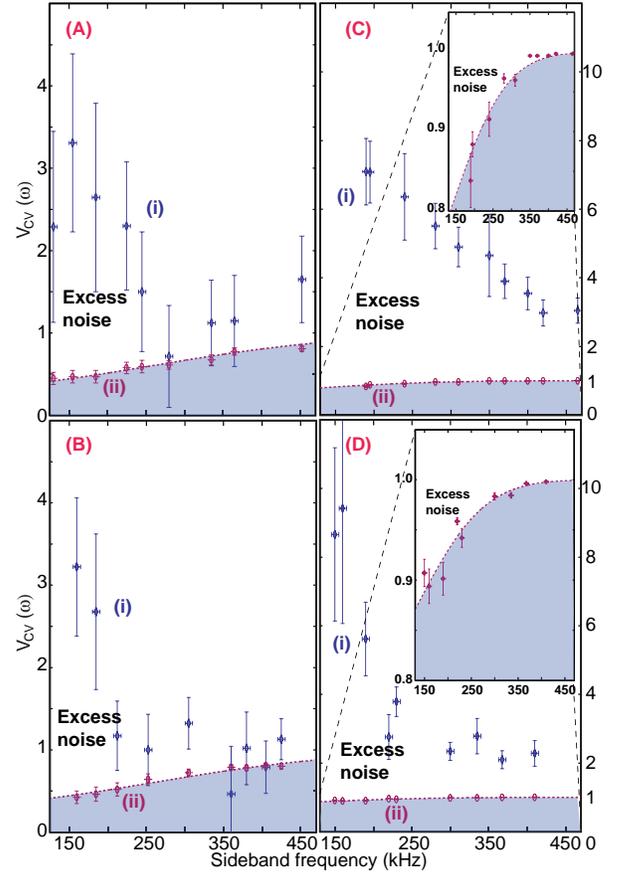}
\caption{$V_{\rm in|out}^{\pm}$ measurements for 2 cell temperatures.  (A) Amplitude quadrature, $42^{\circ}$C; (B) phase quadrature, $42^{\circ}$C; (C) amplitude quadrature, $57^{\circ}$C and (D) phase quadrature, $57^{\circ}$C. The data point groups represent the (i) EIT $V_{\rm in|out}^{\pm}$ and (ii) beamsplitter benchmark $V_{\rm in|out}^{\pm}$. The shape of the shaded area has been fitted using the passive loss described in Eq.~(\ref{abs}). The insets show the zoom-in data points. RBW = 1~kHz, VBW = 30~Hz and 10 averages. The $57^{\circ}$C and $42^{\circ}$C beamsplitter benchmark data points were fitted with $\gamma_0 = 4$~kHz and $\gamma_0 = 3.5$~kHz, respectively. The probe and pump power densities are 9.6~mW/cm$^2$ and 96~mW/cm$^2$, respectively The largest delay for the $57^{\circ}$C and $42^{\circ}$C data is 0.48~$\mu$s and 0.18~$\mu$s, respectively.}
\label{vcv}
\end{center}
\end{figure}
Conditional variance results for two different cell temperatures (corresponding to different atomic densities) are shown in Fig.~\ref {vcv}. The $V_{\rm in|out}^{\pm}$ found using a beamsplitter to simulate the passive loss of the EIT system are the datasets labelled (ii).  Due to the limited bandwidth of EIT, the  passive loss increases with sideband frequency so that in the limit of large frequency, the beamsplitter reference tends to a value of unity.  Using Eq.~(\ref{abs}), the EIT window has been fitted to this beamsplitter data and is represented by the upper limit of the shaded area. The shaded area therefore indicates the area in which a $V_{\rm in|out}^{\pm}$ measurement would be exceeding the quantum measurement limit. EIT data (i) is well above the passive loss benchmark (ii), showing that excess noise is added to the delayed probe beam. Moreover, the excess noise is largest at low frequencies where the passive loss benchmark is at its best.  For higher sideband frequencies the loss in the EIT system dominates the behaviour and $V_{\rm in|out}^{\pm}$ tends to a value of unity.

One source of excess noise is coupling from the pump to the probe since our pump beam has amplitude and phase quadrature noise that lies about 7~dB above the QNL.  By adding amplitude and phase modulation to the pump beam, we were able to measure the transfer functions of the pump-probe coupling.  A maximum coupling of 3~\% for classical phase quadrature signals and 8~\% for classical amplitude quadrature signals, with negligible levels of cross quadrature coupling, was observed. This is only enough to explain 1.2~dB of excess amplitude noise and 0.5~dB of excess phase noise.
 
\begin{figure}[!t]
\begin{center}
\includegraphics[width=7.8cm]{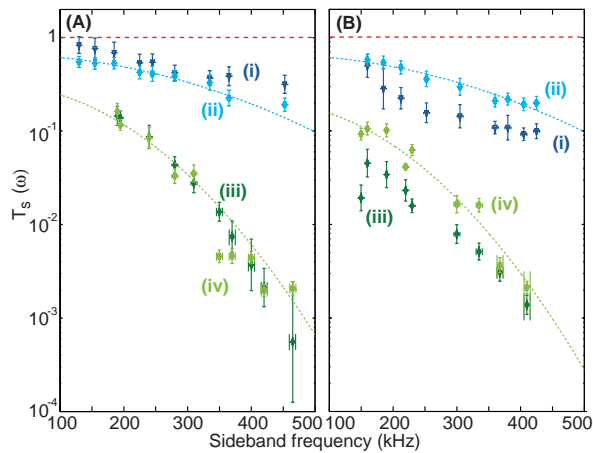}
\caption{Signal transfer coefficient for the (A) amplitude and (B) phase quadratures. (i) $T=42^{\circ}$C and (ii) corresponding beamsplitter benchmark. (iii) $T=57^{\circ}$C and (iv) corresponding beamsplitter benchmark. RBW = 1~kHz, VBW = 30~Hz and 10 averages. The probe and pump power densities are 9.6~mW/cm$^2$ and 96~mW/cm$^2$, respectively. The $57^{\circ}$C and $42^{\circ}$C beamsplitter benchmark data points were fitted with $\gamma_0 = 4$~kHz and $\gamma_0 = 3.5$~kHz, respectively.}
\label{snr}
\end{center}
\end{figure}

We also quantify the performance of our EIT system in terms of the signal transfer between probe input and output.  The signal transfer coefficient is given by $T^{\pm}_{s} (\omega) = {\rm SNR}_{\rm out}^\pm (\omega) / {\rm SNR}_{\rm in}^\pm (\omega)$, where ${\rm SNR}_{\rm out}^\pm (\omega)$ and ${\rm SNR}_{\rm in}^\pm (\omega)$ are the signal-to-noise ratios of the output and input probe fields, respectively. A signal transfer coefficient of unity indicates perfect transfer.   This would be the result for a lossless delay line.  For a passive system with transmission $\eta(\omega)$, the vacuum noise coupled in by the loss gives a signal transfer of $\eta(\omega)$.  Measurements of the signal transfer are shown in Fig.~\ref{snr}. 
The signal transfer degrades as the frequency increases due to the limited bandwidth of EIT.  The EIT system signal transfer is very similar to that of the beamsplitter benchmark indicating that absorption in the EIT system is the dominant cause of reduced signal transfer. There is some deviation from this behaviour for the phase quadrature for both cell temperatures indicating that there is some excess loss of the phase information. 

As discussed above, a deviation of EIT system performance from the passive loss benchmark indicates a discrepancy with the theoretical modelling.  Both in terms of conditional variance and signal transfer we see that the EIT system performance measured in our experiment does not reach the passive loss limit.  The current theoretical modelling \cite{peng} of EIT does not include many effects. In principle, the pump-probe configuration of EIT means that the experiment is performed on atoms of a particular longitudinal velocity class so that any effects of atomic motion can be ignored. Transverse velocities, however, could play a crucial role. The Gaussian intensity profiles of the pump and probe beams mean that atoms with motion in the transverse plane will experience varying optical field intensities, whereas the theory assumes uniform field intensities.  Effects due to high atomic density have also been neglected. Various decoherence mechanisms mean that there is always some fluorescence in the cell.  The probability that these photons are reabsorbed by the atoms grows exponentially with atomic density.  The quantum noise properties of such ``radiation trapping'' \cite{MatskoPRL2001} has not been considered in the context of EIT.  Density dependent effects may be of particular interest since they should be more severe in cold atom systems where the density is higher.  

In summary, our work shows that light delayed by EIT has significant amounts of excess noise that quantum models of EIT are not yet able to explain. This should serve as a motivation for more complete theoretical models to identify the origins of the noise and also as a caveat to claims that EIT in thermal vapour cells is a good method for storing and delaying quantum states. 

The authors thank A.~Peng, W.~P.~Bowen, C.~C.~Harb, J.~J.~Hope and M.~Johnsson for fruitful discussions.  This work was funded by the ARC Centre of Excellence for Quantum-Atom Optics.



\begin{thebibliography}{99}

\bibitem{fleischhauer}{M.~Fleischhauer, and M.~D.~Lukin, Phys.~Rev.~Lett. {\bf 84}, 5094 (2000).}

\bibitem{boller}{K.-J.~Boller, A.~Imamoglu, and S.~E.~Harris, Phys.~Rev.~Lett. {\bf 66}, 2593 (1991).}

\bibitem{eisaman}{M.~D.~Eisaman \text{et al.}, Nature {\bf 438}, 837 (2005).}

\bibitem{chaneliere}{T.~Chaneli\'ere \textit{et al.}, Nature {\bf 438}, 833 (2005).}

\bibitem{kash}{M.~M.~Kash {\it et al.}, Phys.~Rev.~Lett. {\bf 82}, 5229 (1999).}

\bibitem{hau}{L.~V.~Hau, S.~E.~Harris, Z.~Dutton, and C.~H.~Behroozi, Nature {\bf 397}, 594 (1999).}

\bibitem{phillips}{D.~F.~Phillips, A.~Fleischhauer, A.~Mair, R.~L.~Walsworth, and M.~D.~Lukin, Phys.~Rev.~Lett. {\bf 86}, 783 (2001).}

\bibitem{bajcsy}{M.~Bajcsy, A.~S.~Zibrov, and M.~D.~Lukin, Nature {\bf  426}, 638 (2003).}

\bibitem{liu}{C.~Liu, Z.~Dutton, C.~H.~Behroozi, and L.~V.~Hau, Nature {\bf 409}, 490 (2001).}

\bibitem{longdell}{J.~J.~Longdell, E.~Fraval, M.~J.~Sellars, and N.~B.~Manson, Phys.~Rev.~Lett. {\bf 95}, 063601 (2005).}

\bibitem{turukhin}{A.~V.~Turukhin {\it et al.}, Phys.~Rev.~Lett. {\bf 88}, 023602 (2002).}

\bibitem{alexandrov}{E.~B.~Alexandrov, and V.~S.~Zapasskii, Physics-Uspekhi {\bf 47}, 1033 (2004).}

\bibitem{lezama}{A.~M.~Akulshin, A.~Lezama, A.~I.~Sidorov, R.~J.~McLean, and P.~Hannaford, J.~Phys.~B:~At.~Mol.~Opt.~Phys. {\bf 38}, L365 (2005).}

\bibitem{akamatsu}{D.~Akamatsu, K.~Akiba, and M.~Kozuma, Phys.~Rev.~Lett. {\bf 92}, 203602 (2004).}

\bibitem{dantan}{A.~Dantan, and M.~Pinard, Phys.~Rev.~A {\bf 69}, 043810 (2004).}

\bibitem{fleish}{M.~Fleischhauer, and M.~D.~Lukin, Phys.~Rev.~A {\bf 65}, 022314 (2002).}

\bibitem{matsko}{A.~B.~Matsko, Y.~V.~Rostovtsev, O.~Kocharovskaya, A.~S.~Zibrov, and M.~O.~Scully, Phys.~Rev.~A {\bf 64}, 043809 (2001).}

\bibitem{peng}{A.~Peng \textit{et al.},  Phys.~Rev.~A {\bf 71}, 033809 (2004).}

\bibitem{MatskoPRL2001}{A.~B.~Matsko, I.~Novikova, M.~O.~Scully, and G.~R.~Welch, Phys.~Rev.~Lett., {\bf 87}, 133601 (2001).}


\end{thebibliography}
\end{document}